\documentclass[journal,twoside,web]{ieeecolor2}
\usepackage{generic}
\usepackage{cite}
\usepackage{amsmath,amssymb,amsfonts}
\usepackage{algorithmic}
\usepackage[linesnumbered,ruled]{algorithm2e}
\SetKwInOut{Input}{Input}
\SetKwInOut{Output}{Output}
\SetKwInOut{Return}{Return}
\usepackage{graphicx}
\usepackage{textcomp}
\usepackage{array}
\usepackage{multirow}
\usepackage[caption=false,font=normalsize,labelfont=sf,textfont=sf]{subfig}
\usepackage{stfloats}
\usepackage{url}
\usepackage{verbatim}
\usepackage{graphicx}
\usepackage{color}
\usepackage{cite}
\usepackage{xcolor}
\usepackage{mdframed}

\newmdenv[
  linecolor=red,
  linewidth=1pt,
  topline=true,
  bottomline=true,
  rightline=true,
  leftline=true,
  backgroundcolor=red!5,
  skipabove=\baselineskip,
  skipbelow=\baselineskip
]{deletedblock}

\begin{document}
\setlength{\baselineskip}{14pt}

\title{Ray-driven Spectral CT Reconstruction Based on Neural Material Fields}

\author{Ligen Shi, Ping Yang, Chang Liu, Wei Zhang, Xing Zhao*, and Jun Qiu
% <-this % stops a space
\thanks{Corresponding author: Xing Zhao;Ligen Shi and Ping Yang contributed equally to this work.}
\thanks{Ligen Shi is with the College of Computer Science (College of Software), Inner Mongolia University, Hohhot 010021, China (e-mail: ligenshi0826@gmail.com).}% <-this % stops a space
\thanks{Ping Yang is with the Institute of Nuclear and New Energy Technology, Collaborative Innovation Center of Advanced Nuclear Energy Technology, Key Laboratory of Advanced Reactor Engineering and Safety of Ministry of Education, Tsinghua University, Beijing, 100084, China (e-mail: yang\_ping0603@163.com).}
\thanks{Chang Liu and Jun Qiu are with the Institute of Computational Imaging, Beijing Information Science and Technology University, Beijing 100101, China (e-mail: changliuct@gmail.com, qiujun\_sub@126.com).}
\thanks{Wei Zhang is with the National Key Laboratory of Strength and Structural Integrity, Aircraft Strength Research Institute of China, No.86 electronic Second Road, Yanta District, 710065, Xi’an, Shaanxi, China (e-mail: zhangwei\_dut@163.com).}% <-this % stops a space
\thanks{Xing Zhao is with the School of Mathematical Sciences, Capital Normal University, Beijing 100048, China (e-mail: zhaoxing\_1999@126.com).}}

\markboth{Journal of \LaTeX\ Class Files,~Vol.~14, No.~8, August~2023}%
{Shell \MakeLowercase{\textit{et al.}}: A Sample Article Using IEEEtran.cls for IEEE Journals}

\maketitle

\begin{abstract}
\textit{Objective:}
Spectral computed tomography (SCT) material decomposition involves solving an ill-posed inverse problem governed by a nonlinear system of integral equations. To address the resulting instability and limited resolution, we propose Neural Material Fields (NeMFs).
\textit{Methods:}
We model basis materials as continuous vector-valued implicit functions, enabling a neural implicit representation framework for material decomposition. Instead of relying on pixel-driven projection matrices, the NeMFs introduces a lightweight, ray-driven line integral discretization scheme to better approximate physical projections. The neural representation imposes implicit regularization, while a mutual exclusivity regularizer (MER), based on material density distribution differences, promotes better material separation and accelerates convergence. The entire approach is implemented using an automatic differentiation framework.
\textit{Results:}
Experiments demonstrate that NeMFs improves material decomposition accuracy compared to conventional methods. It achieves lower decomposition noise and enables high-resolution reconstruction, mitigating the resolution limitations of standard SCT reconstruction algorithms.
\textit{Conclusion:}
The NeMFs model effectively leverages neural implicit representations and a novel discretization scheme to improve both the stability and resolution of SCT material decomposition. The MER further enhances material separability and convergence speed.
\textit{Significance:}
This study presents a new direction in SCT reconstruction by integrating neural fields and physics-based constraints. The proposed approach supports high-resolution reconstruction and offers a robust alternative to traditional pixel-based methods.
\end{abstract}


\section{Introduction}
\IEEEPARstart{C}{omputed} Tomography (CT) is a widely used imaging technology in modern medical diagnostics and industrial inspection. It is valued for its fast scanning speeds and ability to produce high-fidelity reconstructed images. In practical CT applications, X-ray sources emit photons across a broad energy spectrum. Traditional reconstruction algorithms typically ignore this polychromatic nature, computing a nonlinear average of energy-dependent attenuation coefficients. This averaging process can lead to similar or even identical CT values for substances with low density and high atomic numbers, as well as substances with high density and low atomic numbers. This may also lead to inherent image artefacts in CT images, such as beam hardening and streak artefacts. Various dual-spectrum and multi-spectrum CT imaging modes have been proposed to address these image artefacts and improve CT's ability to identify and quantitatively analyze object compositions.

Spectral Computed Tomography (SCT) uses X-ray scans with multiple spectra or different energy levels within a single spectrum to scan an object. This allows it to take advantage of the varying absorption properties of the object for X-rays at various energies. As a result, it acquires a more extensive and detailed dataset compared to conventional CT. 

Kalender et al.\cite{RN36} introduced the basis material decomposition model for SCT. This model aims to reconstruct density images of materials and generate pseudo-monochromatic images by combining them, effectively reducing artefact effects \cite{RN3}. The challenge of basis material decomposition in SCT involves solving a complex and large-scale nonlinear integral equation system. This system is inherently ill-conditioned due to similarities in the variations of mass attenuation coefficients for different materials with X-ray energy (in keV units). Moreover, the ill-conditioning is exacerbated by the fact that as the radiation passes through highly attenuating substances, there is a higher statistical error in the measurement data of low-energy photons compared to those at higher energies. The ill-conditioning of the equation system amplifies noise during the basis material decomposition process, making it susceptible to noise interference and various artefacts.

Recent studies have shown that Multilayer Perceptrons (MLPs), deep fully connected neural networks, exhibit strong capabilities in representing complex low-dimensional signals \cite{2020NeRF}. Unlike traditional methods that rely on discretely sampled arrays, coordinate-based neural representations provide a continuous alternative that overcomes the limitations of fixed spatial resolution. Meanwhile, their implicit regularization properties help suppress noise when solving inverse problems. In this work, we leverage coordinate-based implicit neural networks to represent base materials, parameterizing their continuous 3D vector-valued functions as Neural Material Fields (NeMFs). To address the forward process of polychromatic projections in SCT, we propose a ray-driven line-integral discretization model based on NeMFs, which eliminates the need for explicitly constructing large sparse system matrices. Furthermore, we introduce a mutual exclusivity regularizer (MER) to enhance decomposition accuracy by enforcing distinct material distributions, which also accelerates network convergence. Finally, by embedding this formulation into a differentiable framework, the material decomposition task is reformulated as an optimization problem over NeMFs, allowing efficient and accurate recovery of continuous material density representations. The main contributions of this study are as follows:
\begin{enumerate}
\item{The proposed ray-driven formulation integrates continuous neural field representations into the polychromatic projection model, thereby avoiding image discretization and promoting a representation that aligns more closely with the continuous nature of X-ray attenuation.}
\item{A mutually exclusive constraint penalty term is introduced to enhance the accuracy of material decomposition. This constraint limits the density distribution differences between different materials, allowing the model to better differentiate material components and improving the convergence speed of the network.}
\item{An optimization model for the nonlinear inverse problem of material decomposition is proposed based on neural fields, under the implicit representation framework for materials. The deep learning automatic differentiation framework is utilized to enable the network to learn the true material density, thereby training a continuous material density model.}
\end{enumerate}

\section{Related Work}
This section reviews representative methods for SCT material decomposition and summarizes recent developments in neural-field-based approaches for tomographic reconstruction.

\subsection{Material Decomposition in SCT Imaging}
Direct image domain decomposition algorithms are based on conventional reconstruction techniques such as the Simultaneous Algebraic Reconstruction Technique (SART) and Filtered Back Projection (FBP), where polychromatic projection data is directly reconstructed into images, and subsequently mapped to material images through inverse transformations. This approach is known as the Brooks method \cite{Brooks}. However, it has limitations as it assumes monochromatic projections and does not account for the polychromatic nature of X-rays. This leads to artefacts such as cupping and streaking \cite{RN38, RN39}. To mitigate these effects, prior corrections or polynomial approximations are typically used \cite{RN40, RN41}.

Projection domain direct mapping methods in SCT convert polychromatic projection data into monochromatic projections corresponding to each material. Separate reconstructions are then performed using traditional algorithms like SART and FBP. Due to the complex interactions between radiation and materials, this approach involves modelling nonlinear functions, spectral characteristics, detector responses, noise models, and energy-dependent mass attenuation coefficients of materials. To simplify the process, researchers often use lookup tables \cite{RN44}, polynomial fittings \cite{RN45, RN46}, or piecewise polynomials \cite{RN47}. While these methods require fewer parameters and are relatively easy to calibrate, their accuracy is inherently limited by the expressiveness of the approximation and discrepancies between calibration phantoms and actual complex objects.

Iterative decomposition algorithms address the challenges of nonlinear mapping in SCT, particularly in scenarios with scanning paths that lack geometric consistency. Among them, E-ART\cite{2014EART}, rectifies polychromatic projections into monochromatic projections of materials and updates the  material images using consistency conditions. This approach effectively tackles projection-domain reconstruction under inconsistencies. Although iterative methods require more computational resources, they offer superior noise suppression capabilities and improved results when combined with physical models. Several subsequent algorithms have been developed to improve convergence rates \cite{ESART, IFBP, pan2023fast}. Notably, IFBP\cite{IFBP}  method employs FBP for residual image reconstruction, which enables a high degree of parallelism and accelerates the convergence process. Specifically, weighted residuals derived from polychromatic projections are first computed and reconstructed via FBP. The corrected material images are then obtained by leveraging pixel-wise correspondences, which effectively mitigates projection inconsistencies and improves decomposition accuracy. 

Beyond the above iterative formulations, a range of optimization-based methods have been developed to address challenges in DECT material decomposition and polyenergetic CT reconstruction. The dynamic nonconvex primal–dual (dNCPD) algorithm of Chen et al.~\cite{Pan2025} integrates a basis-region model with a volume-conservation constraint into a nonconvex optimization framework, providing a representative model-based strategy for stabilizing dual-energy material decomposition. Long and Fessler~\cite{long2014multi} proposed a penalized-likelihood formulation with edge-preserving regularization, solved by a pixel-wise separable quadratic surrogate (PWSQS) algorithm, to suppress noise and cross-talk among basis images. In polyenergetic tomographic imaging, Chung et al.~\cite{chung2010numerical} developed a statistically motivated reconstruction method that models the polyenergetic forward operator to alleviate beam-hardening artifacts. Landi et al.~\cite{landi2015numerical} formulated digital breast tomosynthesis reconstruction as a nonlinear least-squares problem and employed a Levenberg–Marquardt scheme for optimization. Sawatzky et al.~\cite{sawatzky2014proximal} introduced a multi-channel penalized weighted least squares (PWLS) framework that exploits statistical correlations among material-decomposed sinograms and proposed a proximal alternating direction method of multipliers (ADMM) algorithm for efficient minimization. Barber et al.~\cite{barber2016algorithm} presented a constrained one-step inversion method that directly reconstructs basis maps from transmission photon counts using a primal–dual algorithm with a quadratic majorization of the nonconvex data discrepancy. Schmidt et al.~\cite{schmidt2017spectral} further extended this direction with the constrained one-step spectral CT image reconstruction (cOSSCIR) framework, which jointly estimates basis material maps and spectral-response scaling factors from photon-counting detector data while enforcing convex constraints to improve decomposition accuracy and reduce artifacts. Model-based statistical reconstruction has also played a central role in early polyenergetic CT research. Elbakri and Fessler~\cite{elbakri2002statistical} formulated a penalized-likelihood approach that explicitly incorporates the X-ray spectrum and energy-dependent attenuation within a polyenergetic forward model. Their ordered-subsets algorithm estimates voxel-wise material densities under a non-overlapping material assumption and substantially reduces beam-hardening artifacts, providing one of the foundational statistical frameworks for physics-informed reconstruction.

\begin{figure*}[!ht]
\centering
\includegraphics[width=0.90\textwidth]{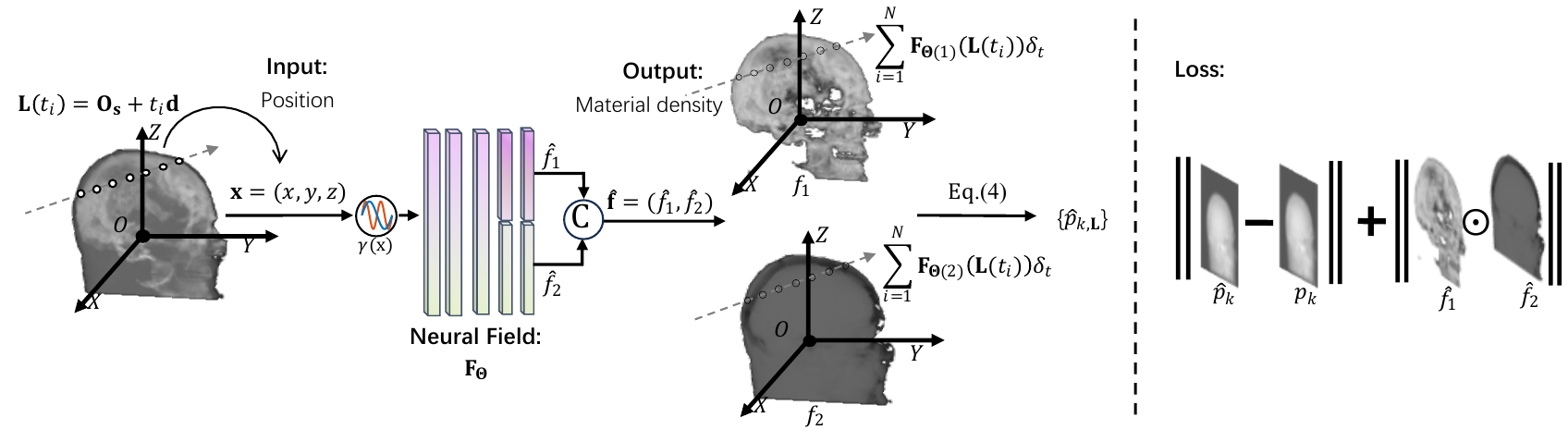}
\caption{Framework of polychromatic projection for dual-material decomposition using NeMFs. Given sampling point coordinates $\mathbf{x} = (x, y, z)$ along a ray $\mathbf{L}(t)$, the network predicts material densities $\hat{\mathbf{f}} = (\hat{f}_1, \hat{f}_2)$. These predictions are integrated to compute the polychromatic projection $\hat{p}_{k,\mathbf{L}}$ under the $k$th spectrum via \eqref{Eq6_}. The loss between $\hat{p}_{k,\mathbf{L}}$ and the measured projection $p_k$ is used for backpropagation.}
\label{fig0}
\end{figure*}

Learning-based methods are increasingly recognized as practical tools for SCT material decomposition. Several studies have successfully employed deep learning frameworks to simulate the complex nonlinear relationship between high and low- spectral projection values and material decomposition coefficients. Various deep learning architectures—including U-Net \cite{materialDecompositionUNet}, GANs \cite{materialDecompositionGAN}, ResNet \cite{RN99}, Butterfly networks \cite{butterflyNet}, and other CNN variants \cite{RN88, RN89, RN98}—have been explored to map dual-spectrum projections or reconstructed images directly to material decomposition coefficients. Although deep learning offers powerful modeling capabilities, its reconstruction accuracy heavily depends on large-scale annotated datasets. However, obtaining such labeled data often requires substantial manual effort and presents challenges in various practical settings.

\subsection{Neural Fields for Tomographic Reconstruction}
Neural fields (NFs), or coordinate-based implicit neural representations, have been applied to a range of tomographic reconstruction problems. NeRF
\cite{2020NeRF} demonstrated that multilayer perceptrons (MLPs), combined with positional encoding, can represent continuous volumetric functions through differentiable rendering, motivating subsequent NF-based studies in medical and computational imaging~\cite{molaei2023implicit}.

A variety of NF approaches have been developed for CT reconstruction. Operating in the projection (sinogram) domain rather than the image domain, CoIL~\cite{COIL} predicts continuous projections from sparse views; IntraTomo~\cite{Intratomo} refines density fields using sinusoidal priors and geometric constraints; NeAT~\cite{Neat} integrates neural feature grids with sparse octrees; NeRP~\cite{NeRP} reconstructs CT or MRI from minimal measurements; DynamicCT~\cite{DynamicCT} models temporal motion; neural attenuation fields (NAF)~\cite{NAF} improves sparse-view CBCT reconstruction; and UncertaINR~\cite{Uncertainr} incorporates uncertainty estimation. Furthermore, the inherent flexibility of NFs allows them to serve as expressive parameterizations for a wide variety of data domains and signal representations. For example, \cite{11049902} represents the sinogram as a continuous neural field and couples it with a differentiable forward model for stripe–artifact suppression, while Liu et al.~\cite{liu2022recovery} embed an NF within an optical forward model to recover continuous 3D refractive-index fields from intensity-only measurements. Beyond reconstruction, implicit neural representations (INRs) can also serve as compact parameterizations for high-dimensional signals, as demonstrated by COIN~\cite{dupont2021coin} for image compression.

While these approaches demonstrate the versatility of NFs, they primarily focus on monochromatic CT or other imaging modalities. SCT, however, requires handling a nonlinear polychromatic forward operator together with the decomposition of multiple interacting basis materials—requirements that differ from those considered in prior NF formulations.

NF architectures also differ in their ability to represent high-frequency
content. Standard MLPs exhibit spectral bias~\cite{SpectralBias}, which limits their capacity to model rapidly varying structures. A variety of architectural designs have been proposed to mitigate this effect, as summarized in recent surveys~\cite{essakine2025where, xie2022neural, molaei2023implicit}. Examples include SIREN~\cite{SIREN}, which uses sinusoidal activations together with the initialization strategy introduced in the original paper, and INCODE~\cite{10483666}, which incorporates an implicit projection-correlation operator to model local consistency among neighboring projections. Positional encoding~\cite{PE} provides a practical alternative by enhancing frequency expressiveness while maintaining stable optimization behavior.

In this work, we adapt the NF framework to SCT material decomposition by introducing two components not addressed in prior NF literature: (i) a ray-driven differentiable \emph{polychromatic} forward model for end-to-end optimization of basis-material densities, and (ii) a mutual exclusivity regularizer (MER) that encourages spatial disjointness among materials to reduce cross-material interference. Together with positional encoding, these components extend the applicability of NFs to the nonlinear mixed-material
setting of spectral CT.

\section{Methods}
This section details the proposed method. Fig. \ref{fig0} illustrates the general workflow of the proposed approach. The decomposed materials are represented as a continuous 3D vector-valued function, parameterized using NeMFs. We then introduce a discretization model for the forward process of material decomposition in polychromatic projections based on the NeMFs. To address the inverse problem, we design an optimization framework that incorporates an MER and a Huber loss function, and solve it using automatic differentiation to recover NeMFs.

\subsection{Parametric Modeling}
In SCT, the measured polychromatic projection data can be modeled using a basis material decomposition framework. Assuming a known X-ray spectrum and negligible scatter, the object’s attenuation is represented as a linear combination of $M$ basis materials with known mass attenuation coefficients $\theta_m(E)$ and spatial distributions $f_m(\mathbf{x})$. The forward model is given by
\begin{equation} 
\begin{gathered} 
p_k(L)=-\ln \int S_k(E)\text{e}^{-\sum_{m=1}^M \theta_m(E) \int_{L} f_m(\mathbf{x}) \mathrm{d}l} \mathrm{d}E, \\
L \in \Pi_k,\quad k=1,2,\cdots,K,
\end{gathered} 
\label{Eq3} 
\end{equation}
where $S_k(E)$ denotes the normalized spectrum of the $k$th acquisition and $\Pi_k$ represents the set of ray paths corresponding to the $k$th spectrum. The objective is to reconstruct the material density maps $f_m(\mathbf{x})$ from the measured projections $\{p_k(L)\}$.

To this end, we represent the density and structural information of the object using $M$ basis materials, and define the corresponding material distributions as a continuous 3D vector-valued implicit function $\mathbf{f}(\mathbf{x})$. We adopt a neural network model $\mathbf{F}_\mathbf{\Theta}$, parameterized by a MLP, to approximate this implicit function—referred to as the Neural Material Fields (NeMFs). The network takes the spatial coordinate $\mathbf{x} = (x, y, z)$ as input and outputs $M$ material densities, i.e., $\mathbf{f}(\mathbf{x}) = \left(f_1(\mathbf{x}), f_2(\mathbf{x}), \cdots, f_M(\mathbf{x})\right)$. The overall mapping is formally expressed as:
\begin{equation}
\mathbf{F}_{\mathbf{\Theta}}: \mathbf{x}\rightarrow \mathbf{f}(\mathbf{x}).
\label{Eq4}
\end{equation}
The weights $\mathbf{\Theta}$ are optimized to ensure that each input 3D coordinate is accurately mapped to its corresponding  material density.

\subsection{Discretization of the Forward Process}
To numerically approximate the continuous forward model in \eqref{Eq3}, we discretize both the X-ray spectrum and the ray path. For the X-ray spectrum, we uniformly divide the energy range into $\Omega_k$ intervals with spacing $\delta_E$, and sample the values of $S_k(E)$ and $\theta_m(E)$ within each interval to achieve spectral discretization. 

Regarding the discretization of the X-ray ray path, we assume that the X-ray source is located at $\mathbf{O}_s$, and the ray travels along a unit direction vector $\mathbf{d}$ from the source to a detector element. Thus, any X-ray path can be described as $\mathbf{L}(t)=\mathbf{O}_s + t\mathbf{d}$. The effective interval $[R-SOD, R+SOD]$ is uniformly divided into $N$ sampling intervals of width $\delta_t$, and the sampling is performed at the start point of each interval. The sampled points are given by:
\begin{equation}
t_i = t_n + \frac{i-1}{N}(2R),
\label{eq7}
\end{equation}
where $i=1,2,\cdots,N$, $SOD$ is the source-to-object distance, and $SDD$ is the source-to-detector distance. $H$ and $W$ denote the detector height and width, respectively. The radius of the effective field of view (FOV) is defined as:
$$
R = \frac{SOD \times 0.5 \min(H, W)}{\sqrt{SDD^2 + (0.5 \min(H, W))^2}}.
$$

This sampling strategy simplifies implementation while maintaining sufficient numerical accuracy for CT reconstruction, based on the assumption of approximate homogeneity within each subinterval. Therefore, the discretized form of \eqref{Eq3} can be obtained by summing the values at $N$ sample points, each multiplied by the interval length $\delta_t$.
\begin{equation} 
	\begin{gathered} 
		p_{k,L}=-\ln \sum_{E=1}^{\Omega_k} S_{k, E} \delta_E \mathrm{e}^{-\sum_{m=1}^M \theta_m(E) \sum_{i=1}^{N}\mathbf{F}_{\mathbf{\Theta}(m)}\left(\mathbf{L}(t_i)\right)\delta_t}, \\ L \in \Pi_k, k=1,2, \cdots, K, \end{gathered} 
	\label{Eq6_} 
\end{equation} 
where $S_{k, E}$ and $\theta_{m,E}$ represent the sampling values of $S_k(E)$ and $\theta_m(E)$ in the $E$th subinterval, respectively. $\mathbf{F}_{\mathbf{\Theta}(m)}$ denotes the $m$th output channel of $\mathbf{F}_{\mathbf{\Theta}}$, corresponding to the density of the $m$th material.

\subsection{Inverse Problem Solving}
The material decomposition process is a noise-amplifying process that is susceptible to noise and various artifacts. Since Huber loss is robust to outliers, we employ Huber loss to optimize the error between the predicted projection $\hat{p}_{k,L}$ and the measured projection $p_{k,L}$ along $L$.
\begin{equation}\label{Huber}
\begin{aligned}
&\text{Huber}\left(\hat{p}_{k,L}, p_{k,L}\right)= 
\left\{
\begin{aligned}
&\frac{1}{2} \left(\hat{p}_{k,L} - p_{k,L}\right)^2, \text{if } \left|\hat{p}_{k,L} - p_{k,L}\right| \leq \delta, \\
&\delta \left|\hat{p}_{k,L} - p_{k,L}\right| - \frac{1}{2} \delta^2, \text{otherwise.}
\end{aligned}
\right.
\end{aligned}
\end{equation}

Although the object is composed of multiple basis materials, their supports are often disjoint, as each location is typically dominated by a single basis material. For instance, in water-bone dual-material decomposition, a high water density at a given position typically results in a near-zero bone density, and vice versa, indicating a mutually exclusive or negatively correlated relationship. To incorporate this property into the optimization process, a mutual exclusivity regularizer (MER) is introduced. The MER adopts a pairwise interaction form, which provides stronger enforcement of mutual exclusivity for multi-material decomposition. The MER is defined as:
\begin{equation}\label{Penalty}
\text{MER}\left(\mathbf{f}\right)=\frac{1}{N} \sum_{i=1}^N \sum_{m=1}^{M-1} \sum_{n=m+1}^{M} f_{m}\left(i\right) \cdot f_{n}\left(i\right),
\end{equation}
where $f_{m}(i)$ denotes the density of the $m$th basis material at the $i$th position. The pairwise formulation penalizes the co-occurrence of any pair of materials at the same spatial location, thereby promoting mutually exclusive material distributions. This formulation ensures that the penalty is nonzero whenever any two materials have nonzero densities at the same voxel, providing effective regularization for multi-material decomposition scenarios.
Crucially, to ensure the mathematical stability of this pairwise penalty and prevent the optimization from diverging towards negative values, a ReLU activation function is applied at the network's output layer. This explicit non-negativity constraint ($f_m(i) \ge 0$) guarantees that the MER penalty is valid and physically meaningful.

Under the MER, we optimize the network parameters $\Theta$ by minimizing the Huber loss between the predicted projection $\hat{p}_{k,L}$ and the measured projection $p_{k,L}$ across all rays:
\begin{equation} \label{Eq8}
\begin{aligned}
{\Theta^*}=&\underset{\Theta}{\operatorname{arg} \operatorname{min}}\frac{1}{|\Pi_k|}\sum_{L \in \Pi_k}\text{Huber}\left(\hat{p}_{k,L}, p_{k,L}\right) \\ &+\alpha\cdot\text{MER}\left(\hat{\mathbf{f}}\right), k=1,2, \cdots, K ,
\end{aligned}
\end{equation} 
where $\Pi_k$ denotes the set of rays under the $k$th X-ray spectrum, $|\Pi_k|$ is the total number of rays, $\hat{\mathbf{f}}$ represents the predicted basis material densities by the network, and $\alpha$ is a regularization hyperparameter. Algorithm \ref{alg_nbmf} describes the procedure for training the NeMFs model.

\begin{algorithm}[!ht]
\caption{Training Process of the NeMFs Model}
\label{alg_nbmf}
\Input{Polychromatic projections $\{p_k\}_{k=1}^K$; mass attenuation coefficients $\{\theta_m(E)\}_{m=1}^M$; X-ray spectra $\{S_k(E)\}_{k=1}^K$; projection angles $\Phi$; geometry ($SOD$, $SDD$, detector size); hyperparameters ($T$, $N$, $\alpha$, $\eta$).}
\Output{Trained NeMFs model $\mathbf{F}_{\Theta}$}
Initialize network parameters $\Theta$;\\ 
Compute field-of-view radius $R$;\\
\For{epoch $t = 1$ to $T$}{
    Randomly select a projection angle $\phi \in \Phi$; \\
    Determine all rays associated with angle $\phi$ and sample points $\mathbf{L}(t_i)$ in $[R-SOD,\; R+SOD]$; \\[2pt]

    Compute predicted material densities 
    $\hat{\mathbf{f}}\big(\mathbf{L}(t_i)\big) = \mathbf{F}_{\Theta}\big(\mathbf{L}(t_i)\big)$; \\

    Compute mutual exclusivity regularizer:
    $\mathrm{MER}\!\left(\hat{\mathbf{f}}\right)$; \\

    \For{$k = 1$ to $K$}{
        Compute predicted projection $\hat{p}_{k,L}$ for spectrum $k$; \\
        Compute data fidelity term $\ell_k = \mathrm{Huber}(\hat{p}_{k,L},\, p_{k,L})$;
    }

    Compute total loss:
    $\ell = \sum_{k=1}^K \ell_k + \alpha \cdot \mathrm{MER}\!\left(\hat{\mathbf{f}}\right)$; \\

    Update parameters using Adam optimizer: 
    $\Theta \leftarrow \mathrm{Adam}(\nabla_{\Theta}\ell, \eta$);
}
\Return{$\mathbf{F}_{\Theta}$;}
\end{algorithm}

\section{Experiments}
This section evaluates the performance of the proposed method for dual-spectral dual-material decomposition using simulated data under several conditions, including noise-free projections, sparse-angle acquisition, and geometric inconsistency with measurement noise. We also include an ablation study to examine the contribution of the MER. Additionally, we present real-data experimental results to further validate the method's performance. Detailed analyses of high-resolution material-density reconstruction, dual-material decomposition under single-spectral information, and three-material decomposition scenarios are provided in the Supplementary Material.

To reduce inverse-crime effects, the simulated projection data include Poisson noise, so the measurements do not match the analytical forward model exactly. This mismatch prevents the network from simply fitting an identical forward operator. In addition, the real-data experiment naturally avoids inverse crime because the physical acquisition process differs from the model assumptions used in reconstruction.

\begin{figure}[!ht]
\centering
\includegraphics[width=0.40\textwidth]{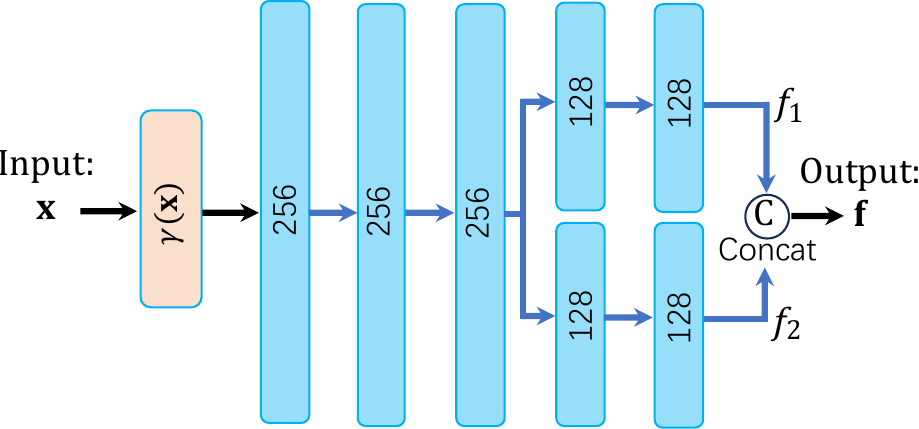}
\caption{The architecture diagram illustrates the fan-beam NeMFs. Hidden layers, depicted in blue, indicate the vector dimensions within each block; blue arrows denote connections to ReLU activation functions. A ReLU activation is applied at the output layer to enforce non-negativity of material densities.}
\label{fig4}
\end{figure}

For simplicity, all experiments are conducted using fan-beam CT. The network architecture is shown in Fig.~\ref{fig4}. Quantitative evaluation is performed using peak signal-to-noise ratio (PSNR) and structural similarity index (SSIM). We parameterize the neural field using an MLP with positional encoding, which provides stable optimization and sufficient frequency representation for the material-density field in our formulation. Formally, $\mathbf{F}_\mathbf{\Theta}$ is defined as a composition $\mathbf{F}_\mathbf{\Theta} = \mathbf{F}_\mathbf{\Theta}^{\prime} \circ \gamma$, where $\mathbf{F}_\mathbf{\Theta}^{\prime}$ is a standard MLP, and $\gamma$ denotes the Fourier feature mapping given by:
\begin{equation}  
\begin{aligned}
\gamma(p) = &(\sin(2^0 \pi p), \cos(2^0 \pi p), \cdots, \\
&\sin(2^{D-1} \pi p), \cos(2^{D-1} \pi p)),
\end{aligned}
\label{eq4.4} 
\end{equation}
where $\gamma(\cdot)$ is applied to each coordinate of the input $\mathbf{x} \in [-1, 1]$, and $D$ denotes the encoding dimensionality. The network has approximately 0.21 M parameters.

\subsection{Dual-Spectral Dual-Material Experiments}
This section uses dual-spectrum X-ray scanning to reconstruct dual-material images, where $M=K=2$. Both the E-ART and IFBP belong to iterative methods, exhibiting robust noise resistance and are suitable for scenarios with geometric inconsistencies in the scanning paths and different spectra. In addition, the dNCPD algorithm is included for comparison. As a representative model-based method, dNCPD addresses the nonlinear polychromatic CT forward model via non-convex optimization and has demonstrated strong numerical accuracy and stability in multi-basis reconstruction. Therefore, the proposed method is compared with E-ART, IFBP, and dNCPD to comprehensively validate its performance.

\begin{figure}[!ht] 
\centering 
\includegraphics[width=0.40\textwidth]{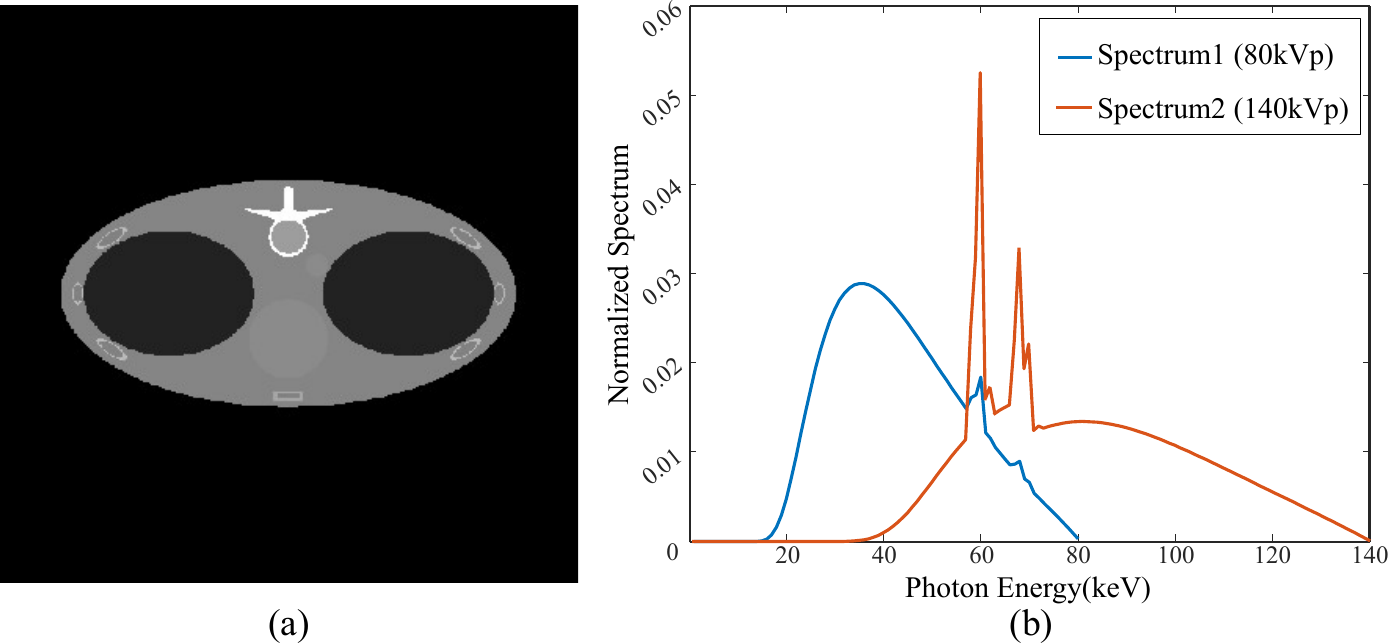} 
\caption{(a) Slice of the thorax phantom used in the numerical simulation; (b) X-ray spectra used in the numerical simulation.} 
\label{fig2} 
\end{figure}

\textbf{Experiment Setups.}\label{Simulated Experiment Scan Configurations}
The experiments use slice data from the 3D FORBILD thorax phantom\cite{forbild_thorax_phantom}. Fig. ~\ref{fig2}(a) illustrates the thorax phantom with a resolution of $512\times 512$. The thorax phantom is composed of materials with varying densities, such as water and bone. Water and bone are also used as the  materials for image reconstruction. The mass attenuation coefficients of the materials (water and bone) are obtained from NIST\cite{nist_xray_mass_attenuation}. The X-ray spectra used in the experiment were simulated using the free software SpectrumGUI\cite{spectrumgui}. The X-ray source is the GE Maxiray 125 X-ray tube, with voltages of 80 kV and 140 kV, the latter filtered by 1mm copper. Fig. \ref{fig2}(b) illustrates the normalized X-ray spectra.

\textbf{Scanning configuration.}
The distance between the X-ray source and the rotation center (Turntable axis) is $SOD = 1000$ mm, and the distance between the source and detector is $SDD = 1536$ mm. The linear detector consists of 512 elements of size $0.8$ mm. Polychromatic projections are generated using the forward model in~\eqref{Eq3}. Based on this geometry, the resulting FOV radius is approximately 132.16 mm. The X-ray spectra and mass attenuation coefficients are sampled at 1 keV intervals.

\textbf{Acquisition setup.}
Unless otherwise specified, all simulated experiments assume dual-energy acquisition in which each projection view is measured at both energy levels. Projection angles are uniformly sampled over a full $360^\circ$ rotation. Three acquisition settings are considered: (1) \emph{Noise-free experiment}: 720 uniformly spaced views per spectrum. (2) \emph{Sparse-angle experiment}: 120 uniformly spaced views per spectrum. (3) \emph{Geometric inconsistency experiment}: 1441 projections are generated per spectrum, and 720 views are selected by taking every other angle. The two energy levels use interleaved angle sets, resulting in controlled angular misalignment between spectra.

\textbf{Algorithm parameters.}
In the dual-spectral dual-material decomposition experiments, the E-ART and IFBP methods are iterated until the relative change between successive iterations falls below $10^{-4}$. IFBP uses a Hamming window for frequency-domain filtering. For the dNCPD method, we adopt the parameter settings recommended in the original paper~\cite{Pan2025}; detailed parameter configurations are provided in the Supplementary Material. We train the network using the Adam optimizer with learning rate $\eta = 1 \times 10^{-3}$. Training is performed for 300 epochs with $D = 8$, $N = 1023$, $\alpha = 10^{-2}$, and $\delta = 1$. All experiments are conducted on an NVIDIA GeForce RTX 3080 GPU.

\subsubsection{Comparison Experiment with Noise-free Data} 
\begin{figure}[!ht]
\centering
\includegraphics[width=0.48\textwidth]{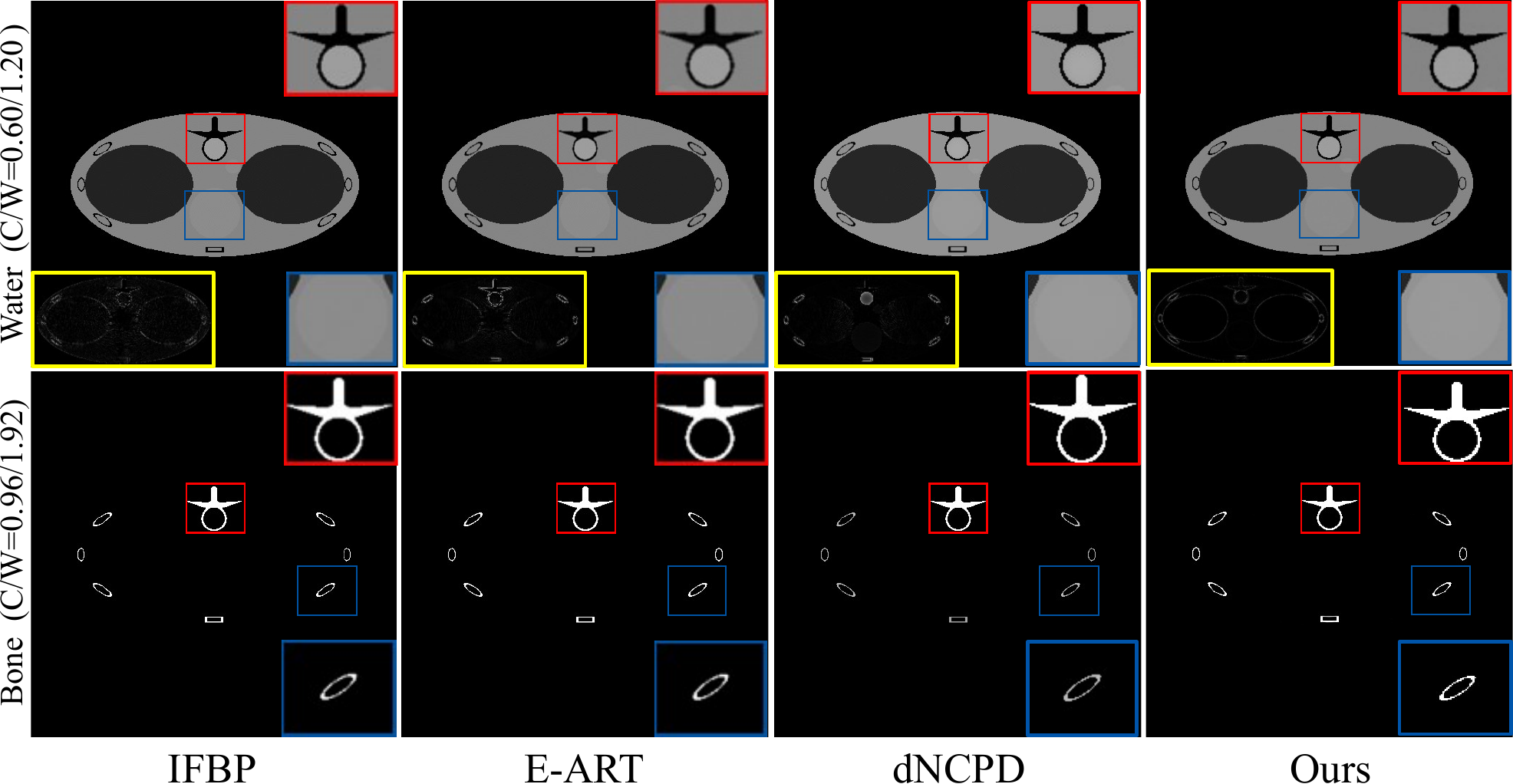}
\caption{Comparison of noise-free basis-material reconstructions obtained using IFBP, E-ART, dNCPD, and the proposed method. For each basis material, the reconstructed image and two local magnified views are shown. For each image in the first row, the lower-left inset (yellow dashed box) presents the residual image with a display window of [0,0.2].}
\label{fig5}
\end{figure}

\begin{figure}[!ht]
\centering
\includegraphics[width=0.50\textwidth]{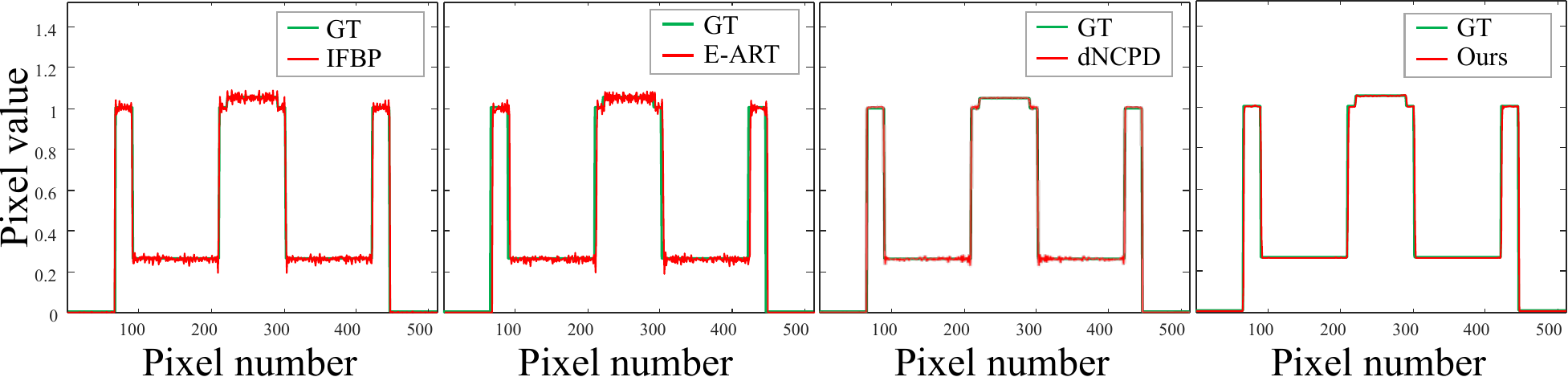}
\caption{Profiles at line 290: Water material density image on noise-free data.}
\label{fig5_Profiles}
\end{figure}

In the noise-free comparison experiment, 720 projections were collected for each spectrum. Fig. \ref{fig5} presents the reconstruction results of the compared algorithms and the proposed method using noise-free data. It can be observed from the images that all four methods achieve satisfactory reconstruction results, with correct material separation. Notably, significant differences exist in the density images of water, as demonstrated by the profile plot along the 290th row in Fig. \ref{fig5_Profiles}. From Fig. \ref{fig5_Profiles}, it is evident that the profile plot of the proposed method is closer to the Ground Truth. Table \ref{tab1} reports the quantitative evaluation results in terms of PSNR and SSIM. While dNCPD achieves a higher PSNR for the high-contrast bone structure, the proposed NeMFs demonstrates a significant advantage in reconstructing the low-contrast water component, outperforming dNCPD by 0.79 dB. This indicates that our method effectively preserves soft-tissue fidelity and suppresses cross-material interference.

\begin{table}[!ht]
\centering
\caption{Quantitative comparison on noise-free data.}
\label{tab1}
\footnotesize
\setlength{\tabcolsep}{2pt}
\renewcommand{\arraystretch}{1.2}
\begin{tabular}{c|cc|cc|cc|cc}
\hline\hline 
\multirow{2}{*}{Material} & \multicolumn{2}{c|}{IFBP} & \multicolumn{2}{c|}{E-ART} & \multicolumn{2}{c|}{dNCPD} & \multicolumn{2}{c}{Ours} \\
\cline {2-9} 
& PSNR & SSIM & PSNR & SSIM & PSNR & SSIM & PSNR & SSIM \\
\hline 
Bone & 43.99 & 0.856 & 35.40 & 0.999 & 52.97 & 0.999 & 45.43 & 0.999 \\
Water & 38.93 & 0.800 & 31.97 & 0.975 & 38.68 & 0.992 & 39.47 & 0.998 \\
\hline \hline
\end{tabular}
\end{table}

\subsubsection{Comparison Experiment with Sparse Angle Data}
\begin{figure}[!ht]
\centering
\includegraphics[width=0.48\textwidth]{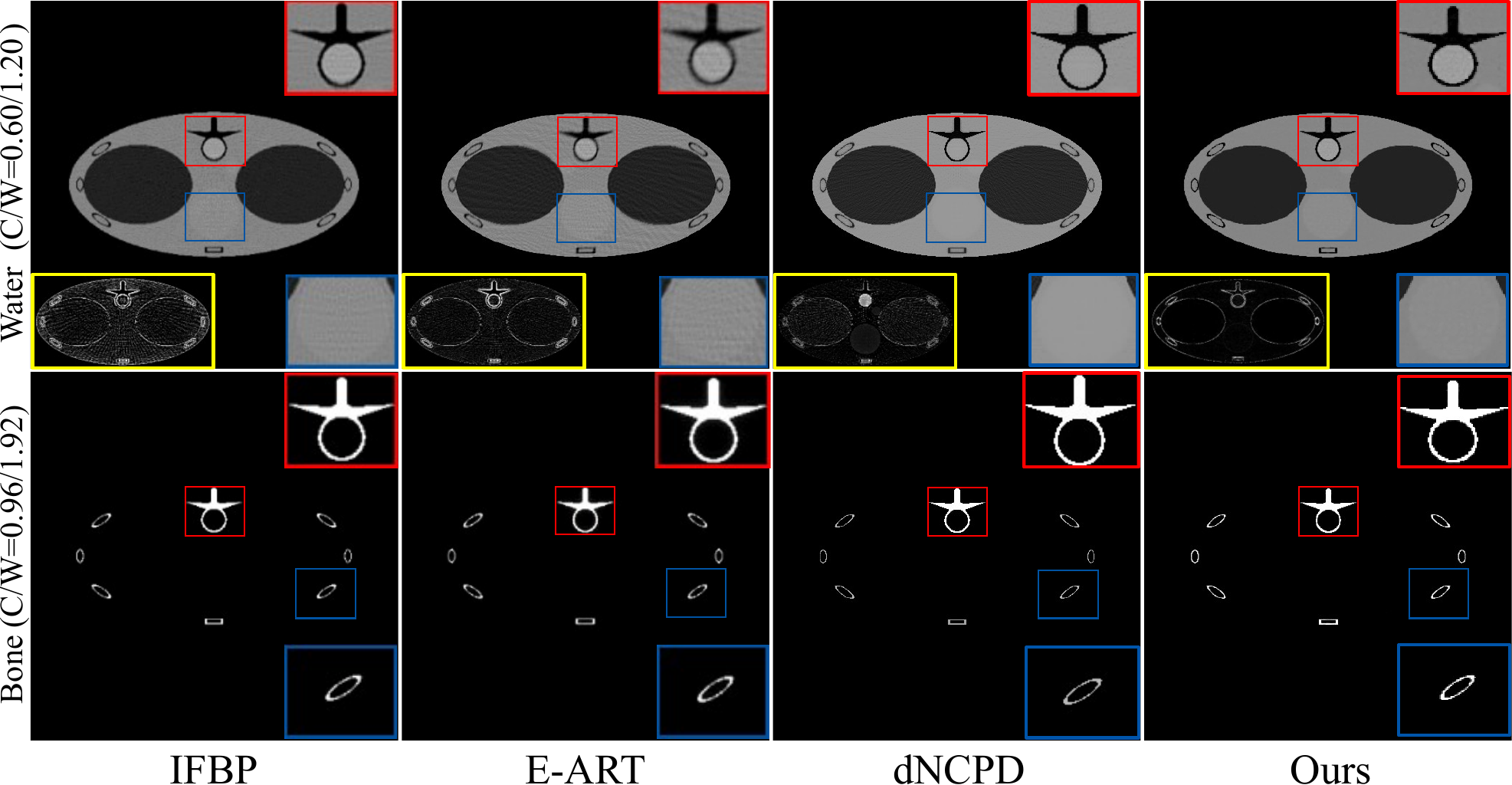}
\caption{Comparison of sparse-angle basis-material reconstruction results obtained by 
IFBP, E-ART, dNCPD, and the proposed method. The layout follows that of Fig.\ref{fig5}.}
\label{fig9_Sparse_angle}
\end{figure}

\begin{figure}[!ht]
\centering
\includegraphics[width=0.50\textwidth]{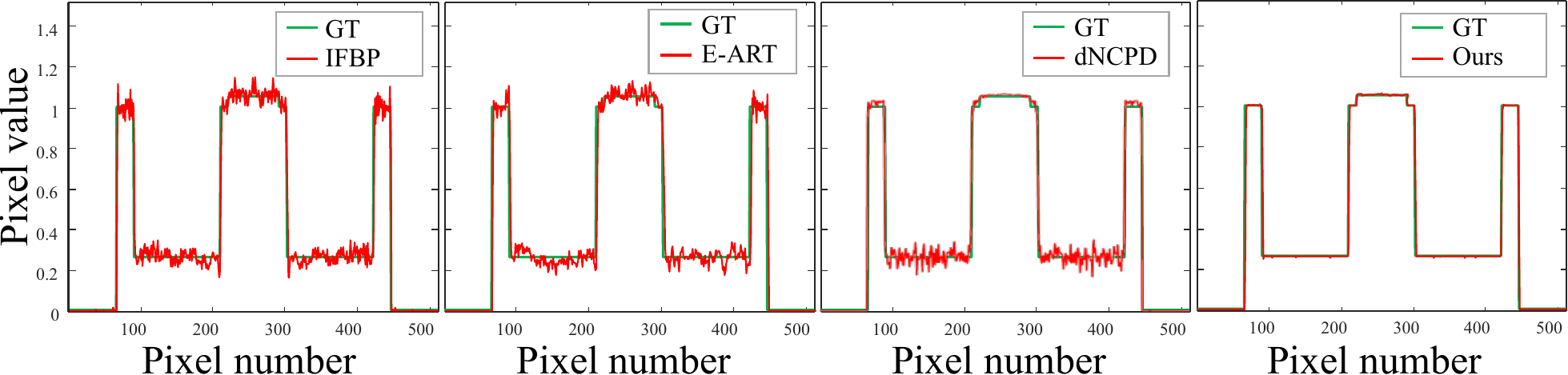}
\caption{Profiles at line 290: Water material density image on sparse angle data.}
\label{fig9_Sparse_angle_Profiles}
\end{figure}

In this experiment, a comparison was conducted using sparse angle spectral data, with 120 projections collected for each spectrum. Fig. \ref{fig9_Sparse_angle} illustrates the reconstruction results of different algorithms using sparse angle data. Observation of Fig. \ref{fig9_Sparse_angle} reveals that IFBP and E-ART exhibit some radial artefacts in the water density image, which are challenging to eliminate even with more iterations. The dNCPD method also presents slight radial artefacts, although to a lesser extent. In contrast, the proposed method shows no apparent artefacts in the reconstructed images. Significant differences are observed in the water density image; hence, the profile plot along the 290th row is provided in Fig. \ref{fig9_Sparse_angle_Profiles} for the compared algorithms and the proposed method. It can be seen from Fig \ref{fig9_Sparse_angle_Profiles} that the proposed method yields a profile plot closer to the Ground Truth for the water density image along the 290th row. Table \ref{tab1_sparse_angle} lists the proposed method's quantitative metrics, including PSNR and SSIM, and the compared algorithms. While dNCPD achieves a higher PSNR for bone, the proposed NeMFs demonstrates substantial advantages in reconstructing the water component, outperforming dNCPD by 2.36 dB in PSNR. This highlights the robustness of our method in challenging sparse-angle scenarios, where the continuous neural representation and MER effectively suppress artifacts and preserve soft-tissue fidelity.

\begin{table}[!ht]
\centering
\caption{Quantitative comparison on sparse-angle data.}
\label{tab1_sparse_angle}
\footnotesize
\setlength{\tabcolsep}{2pt}
\renewcommand{\arraystretch}{1.2}
\begin{tabular}{c|cc|cc|cc|cc}
\hline\hline 
\multirow{2}{*}{Material} & \multicolumn{2}{c|}{IFBP} & \multicolumn{2}{c|}{E-ART} & \multicolumn{2}{c|}{dNCPD} & \multicolumn{2}{c}{Ours} \\
\cline {2-9} 
& PSNR & SSIM & PSNR & SSIM & PSNR & SSIM & PSNR & SSIM \\
\hline 
Bone & 32.79 & 0.462 & 40.55 & 0.998 & 44.27 & 0.998 & 40.22 & 0.995 \\
Water & 28.50 & 0.398 & 28.24 & 0.916 & 32.27 & 0.939 & 34.63 & 0.991 \\
\hline \hline
\end{tabular}
\end{table}

\subsubsection{Comparison Experiment with Noise and Geometric Inconsistency data} 
This subsection evaluates the robustness of the proposed method to noise and geometric inconsistency in SCT data acquisition. To simulate measurement noise, Poisson noise with an initial photon count of $10^6$ was added to the noiseless data to generate noisy multi-spectral projections. Geometric inconsistency was simulated by acquiring 1441 projections with spectrum-dependent angular sampling. Specifically, 720 projections were collected for the 80 kVp spectrum at every other angle starting from 0 angle, while 720 projections were collected for the 140 kVp spectrum at every other angle starting from 1 angle.

\begin{figure}[!ht]
\centering
\includegraphics[width=0.48\textwidth]{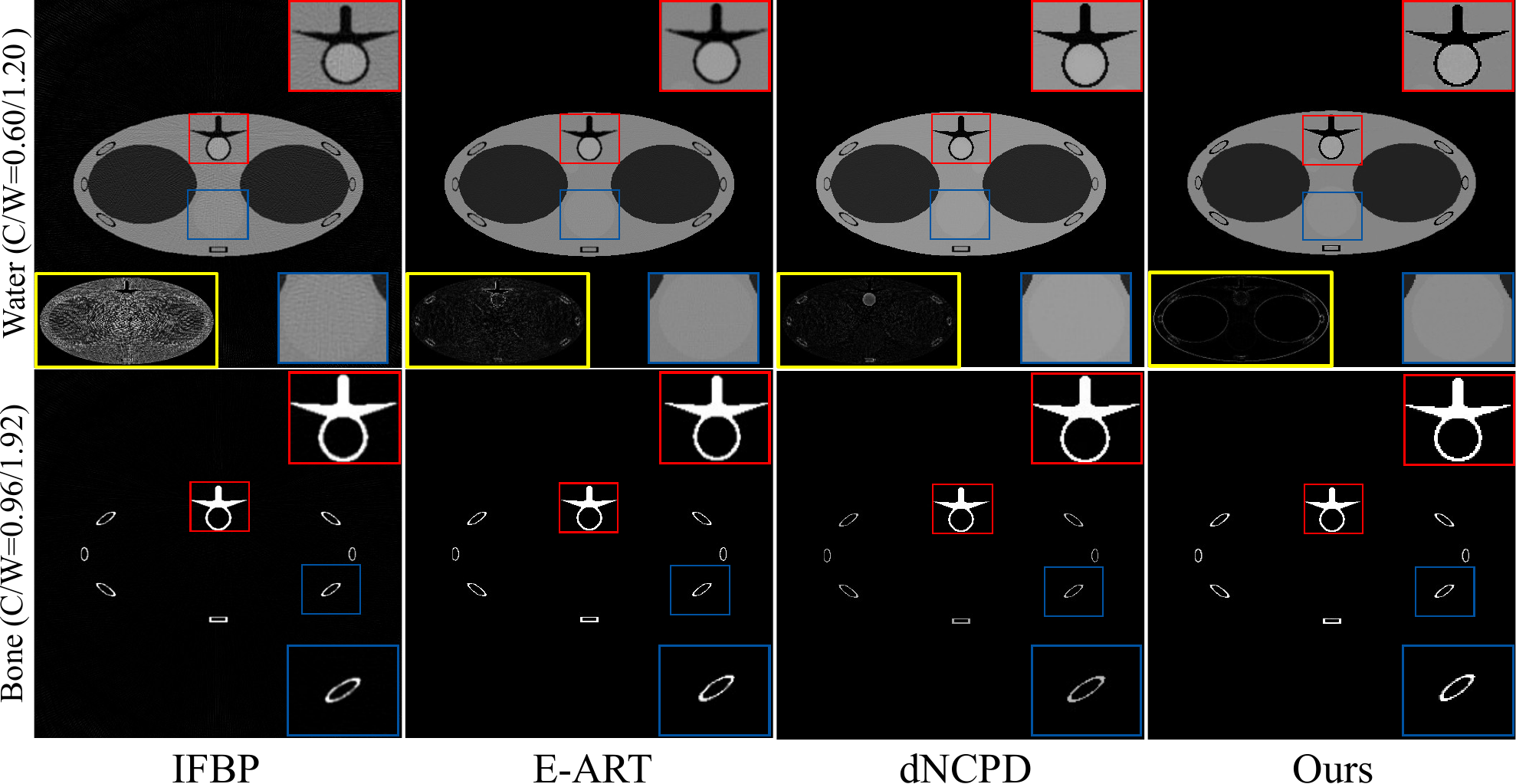}
\caption{Comparison of noise and geometric inconsistency basis-material reconstructions obtained by IFBP, E-ART, dNCPD and the proposed method. The layout follows that of Fig.\ref{fig5}.}
\label{fig7}
\end{figure}

\begin{figure}[!ht]
\centering
\includegraphics[width=0.48\textwidth]{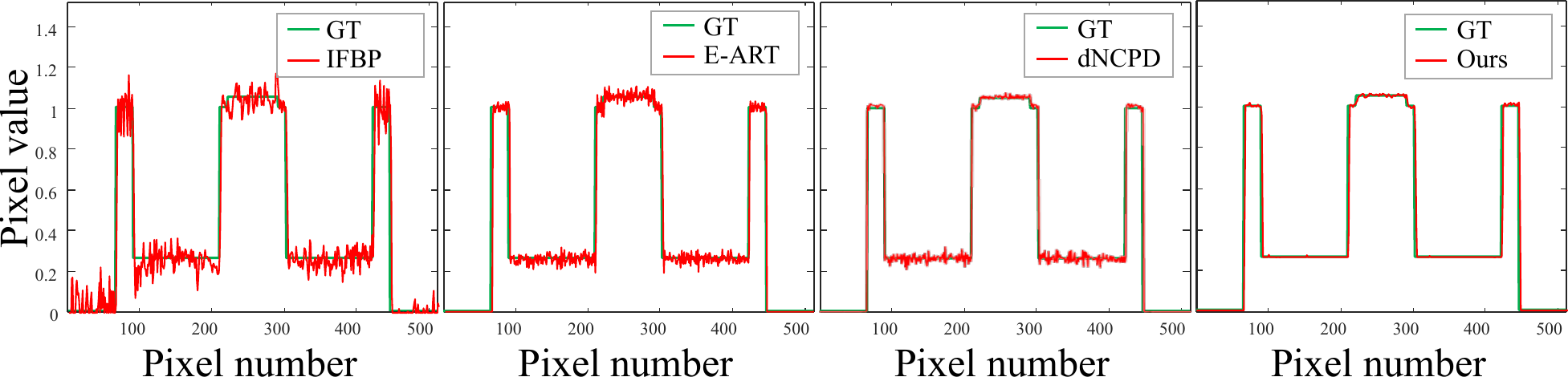}
\caption{Profiles at line 290: Water material density image with noise and geometric inconsistencies.}
\label{fig7_noise_geo_inconsistency_Profiles}
\end{figure}

Fig. \ref{fig7} illustrates the reconstruction results of the compared algorithms and the proposed method with Poisson noise and geometrically inconsistent data. Upon observing the reconstruction results, all four methods achieve good reconstruction performance. However, the proposed method exhibits fewer artefacts in the water material density image. Fig. \ref{fig7_noise_geo_inconsistency_Profiles} provides profile plots along the 290th row for both the compared algorithms and the proposed method in the water material density image. From the profile plots, it can be seen that the proposed method yields smoother material image surfaces, which are closer to the Ground Truth. Table \ref{tab3} lists the quantitative metrics, including PSNR and SSIM. While dNCPD achieves a higher PSNR for bone, the proposed NeMFs demonstrates substantial advantages in reconstructing the water component, outperforming dNCPD by 3.27 dB in PSNR. This indicates that our method is highly robust against noise and geometric inconsistencies, effectively preserving soft-tissue fidelity under challenging acquisition conditions.

\begin{table}[!ht]
\centering
\caption{Quantitative comparison on noisy and geometrically inconsistent data.}
\label{tab3}
\footnotesize
\setlength{\tabcolsep}{2pt}
\renewcommand{\arraystretch}{1.2}
\begin{tabular}{c|cc|cc|cc|cc}
\hline\hline 
\multirow{2}{*}{Material} & \multicolumn{2}{c|}{IFBP} & \multicolumn{2}{c|}{E-ART} & \multicolumn{2}{c|}{dNCPD} & \multicolumn{2}{c}{Ours} \\
\cline {2-9} 
& PSNR & SSIM & PSNR & SSIM & PSNR & SSIM & PSNR & SSIM \\
\hline 
Bone  & 35.72 & 0.782 & 35.26 & 0.998 & 43.82 & 0.998 & 38.13 & 0.998 \\
Water & 27.51 & 0.418 & 30.87 & 0.936 & 30.60 & 0.954 & 33.87 & 0.993 \\
\hline \hline
\end{tabular}
\end{table}

\subsection{Real-Data Experimental Results} 
\begin{figure}[!ht] 
\centering \includegraphics[width=0.5\textwidth]{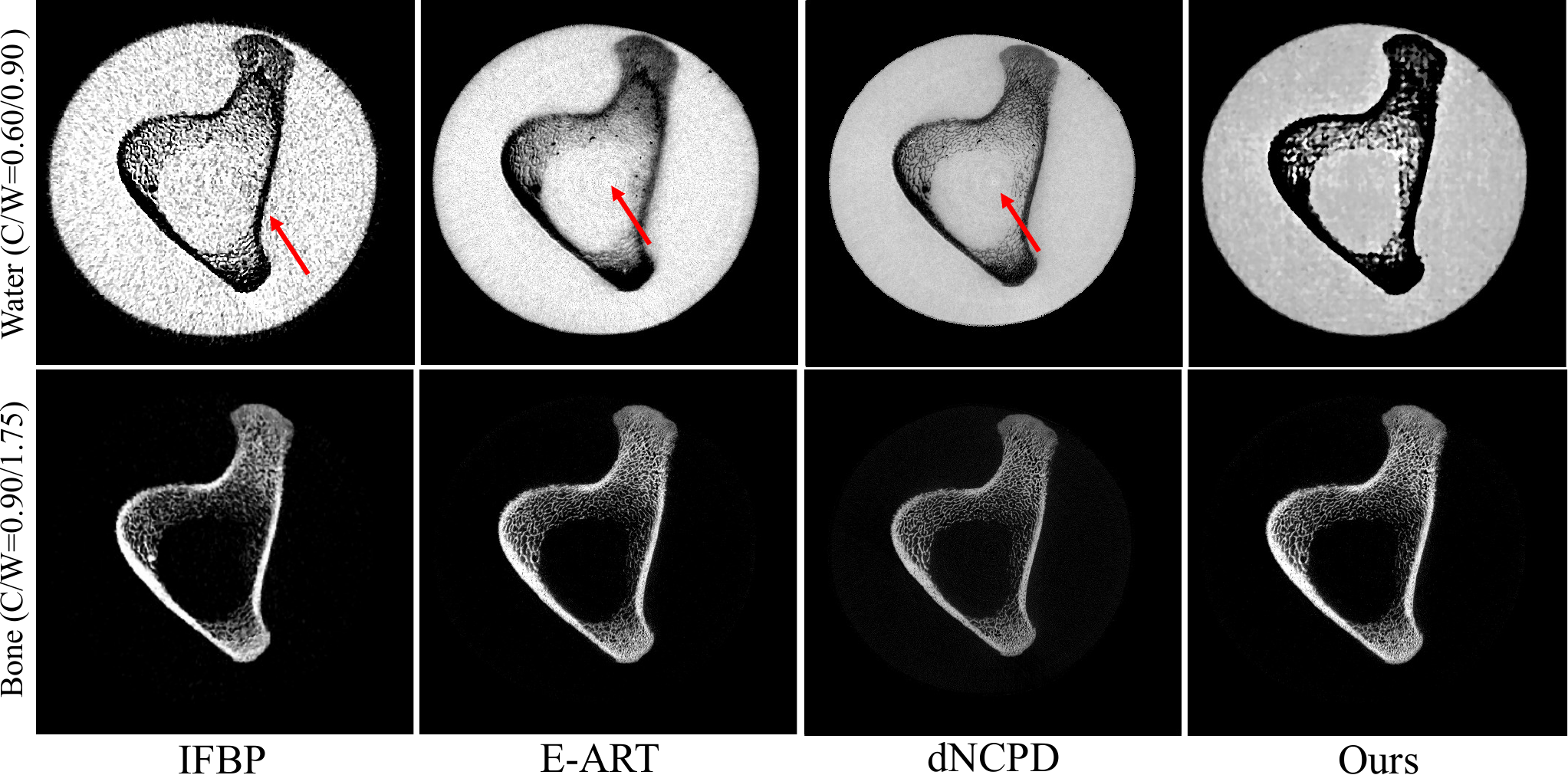}
\caption{Comparison of basis material density image reconstructed by different methods.} \label{NBMF_fig_10_real} 
\end{figure}

This section presents a comparison experiment using real data, with scanning configurations provided in Table \ref{NBMF_tab_6}. The experimental parameters are consistent with those used for the simulated data. Fig. \ref{NBMF_fig_10_real} shows the reconstruction results from the IFBP, E-ART, dNCPD, and proposed methods. The images in Fig. \ref{NBMF_fig_10_real} demonstrate that all four methods produce acceptable water and bone material density images. However, the water material image reconstructed by IFBP exhibits significant noise, and both IFBP, E-ART, and dNCPD show noticeable ring artifacts (indicated by the red arrows) in the water material density image. In contrast, the proposed method does not exhibit such artifacts and produces locally smoother water material density image.

\begin{table}[!ht] 
\renewcommand{\arraystretch}{1.25} 
\begin{center} \caption{Dual-energy CT parameters for real data} \label{NBMF_tab_6} 
\setlength{\tabcolsep}{12pt}
\begin{tabular}{l|cc} 
\hline\hline Scan Parameters & {80 kVp} & {140 kVp} \\
\hline 
Tube Voltage (kVp) & {80} & {140} \\
\hline Tube Current (uA) & {240} & {120} \\
\hline Filter (mm) & {Al 1.5} & {Cu 0.5} \\
\hline SOD (mm) & {355.61} & {355.61} \\
\hline SDD (mm) & {673.96} & {673.96} \\
\hline Number of Scan Angles & 1440 & 1440 \\
\hline \hline 
\end{tabular} 
\end{center} 
\end{table}

\subsection{MER Ablation Experiments}
\begin{table}[!ht]
\centering
\caption{Ablation study: Quantitative comparison with and without MER.}
\label{tab_ablation}
\footnotesize
\setlength{\tabcolsep}{3pt}
\renewcommand{\arraystretch}{1.2}
\begin{tabular}{c|c|cc|cc}
\hline\hline 
\multirow{2}{*}{Setting} & \multirow{2}{*}{Material} & \multicolumn{2}{c|}{w/o MER} & \multicolumn{2}{c}{w/ MER} \\
\cline {3-6} 
& & PSNR & SSIM & PSNR & SSIM \\
\hline 
\multirow{2}{*}{Noise-free} & Bone & 44.99 & 0.999 & \textbf{45.43} & \textbf{0.999} \\
& Water & 39.11 & 0.998 & \textbf{39.47} & 0.998 \\
\hline 
\multirow{2}{*}{Sparse-angle} & Bone & 39.44 & 0.991 & \textbf{40.22} & \textbf{0.995} \\
& Water & 33.04 & 0.987 & \textbf{34.63} & \textbf{0.991} \\
\hline 
\multirow{2}{*}{Noise + Geo. Inc.} & Bone & 37.41 & 0.997 & \textbf{38.13} & \textbf{0.998} \\
& Water & 33.14 & 0.991 & \textbf{33.87} & \textbf{0.993} \\
\hline \hline
\end{tabular}
\end{table}

Table~\ref{tab_ablation} provides quantitative comparison under three experimental settings. The results demonstrate that MER consistently improves both PSNR and SSIM metrics across all three experimental conditions (noise-free, sparse-angle, and noise and geometric inconsistency), which fully validates the effectiveness of the MER regularizer in enhancing material decomposition accuracy. Specifically, the use of MER leads to: \\ 1) Higher PSNR values for both bone and water materials in all scenarios, indicating reduced reconstruction error; \\ 2) Improved SSIM values, reflecting better structural similarity to the ground truth; \\ 3) Particularly significant gains in the sparse-angle scenario, where MER provides up to 1.59 dB improvement in PSNR for water and 0.48 improvement in SSIM for bone, demonstrating its robustness under challenging acquisition conditions. The convergence curve of the training process is provided in the Supplementary Material.

\section{Discussions}
This study presents a novel approach to SCT material decomposition by modeling materials as continuous 3D implicit functions of spatial coordinates, parameterized by fully connected neural networks. Numerical experiments under various conditions—including noise-free data, sparse-angle sampling, Poisson noise, and geometric inconsistency—demonstrate that the proposed method outperforms traditional methods such as IFBP, E-ART, and dNCPD, particularly under sparse-angle and noisy settings, thereby indicating strong robustness for practical applications.

Beyond quantitative accuracy, the proposed NeMF framework offers several distinctive advantages over discrete iterative methods such as dNCPD. First, \textbf{resolution independence}: NeMF represents material density as a continuous function, enabling reconstruction and querying at arbitrary resolutions without retraining or interpolation. As demonstrated in \textbf{the Supplementary Material}, the same trained network can output images at $256 \times 256$, $512 \times 512$, or $1024 \times 1024$ with consistent quality, whereas voxel-based methods are inherently tied to a fixed grid. Second, \textbf{memory efficiency}: the network contains only approximately 0.21 M parameters, which is significantly smaller than storing multiple high-resolution material images. Third, \textbf{continuous representation}: the neural field provides a smooth, differentiable representation of material density that better aligns with the continuous nature of X-ray attenuation, potentially reducing discretization artifacts. These properties make NeMF particularly suitable for applications requiring multi-scale visualization or integration with downstream tasks that benefit from continuous representations.

The method features a lightweight architecture with relatively few parameters and uses a self-supervised optimization scheme that does not require external training datasets. Nonetheless, training the network still involves a trade-off between computation time and memory due to the auto-differentiation framework.

Finally, we note that the method adopts a deterministic fixed-frequency
positional encoding. Although random Fourier feature variants are compatible with our framework and may provide alternative frequency characteristics, they are not investigated in this study. Exploring different encoding schemes may be a useful direction for future work.

\section{Conclusion}
This work introduced a neural-field formulation for spectral CT material decomposition, in which each basis material is represented as a continuous implicit function parameterized by a compact multilayer perceptron. The proposed framework performs reconstruction in a self-supervised manner using polychromatic measurements and does not rely on discretized image grids or external training datasets. By coupling the neural representation with a ray-driven polychromatic forward model and a mutual-exclusivity regularizer, the method enables stable decomposition of dual-energy data under a variety of acquisition conditions. Comprehensive experiments demonstrate that the method achieves robust and high-fidelity decomposition across noise-free, sparse-angle, noisy, and geometrically inconsistent acquisition conditions. These results highlight the potential of neural-field parameterizations to offer a flexible alternative to traditional voxel-based approaches in spectral CT. Future work may further extend the framework by incorporating more detailed measurement physics and adapting it to broader acquisition scenarios.

\bibliography{ref.bib}
\bibliographystyle{IEEEtran} 
\end{document}